\documentclass[]{raa}            

\usepackage{graphicx,times}             
\usepackage{natbib}
\usepackage{amssymb,amsmath}
\bibpunct{(}{)}{;}{a}{}{,}

\usepackage[pagebackref=true]{hyperref}

\begin{document}

  \title{Constraining Brans-Dicke cosmology with the CSST galaxy clustering spectroscopic survey
}

   \volnopage{Vol.0 (20xx) No.0, 000--000}      
   \setcounter{page}{1}          

   \author{Anda Chen
      \inst{1,2}
   \and Yan Gong*
      \inst{1,3}
   \and Fengquan Wu
      \inst{4}
   \and Yougang Wang
      \inst{4}
   \and Xuelei Chen
      \inst{4,2,5}
   }

    \institute{Key Laboratory of Space Astronomy and Technology, National Astronomical Observatories,Chinese Academy of Sciences, Beijing 100101, China; {\it Email: gongyan@bao.ac.cn}\\
        \and
            School of Astronomy and Space Sciences, University of Chinese Academy of Sciences, Beijing 100049, China;\\
        \and
             Science Center for China Space Station Telescope, National Astronomical Observatories, Chinese Academy of Sciences, 20A Datun Road, Beijing 100101, China \\
        \and
             Key Laboratory of Computational Astrophysics, National Astronomical Observatories, Chinese Academy of Sciences, Beijing 100101, China\\
        \and
             Center for High Energy Physics, Peking University, Beijing 100871, China\\
\vs\no
   {\small Received~~20xx month day; accepted~~20xx~~month day}
}

\abstract{ The Brans-Dicke (BD) theory is the simplest Scalar-Tensor theory of gravity, which can be considered as a candidate of modified Einstein's theory of general relativity. In this work, we forecast the constraints on BD theory in the CSST galaxy clustering spectroscopic survey with a magnitude limit $\sim 23$ AB mag for point-source 5$\sigma$ detection. We generate mock data based on the zCOSMOS catalog and consider the observational and instrumental effects of the CSST spectroscopic survey. We predicate galaxy power spectra in the BD theory from $z=0$ to 1.5, and the galaxy bias and other systematical parameters are also included. The Markov Chain Monte Carlo (MCMC) technique is employed to find the best-fits and probability distributions of the cosmological and systematical parameters. A  Brans-Dicke parameter $\zeta$ is introduced, which satisfies $\zeta=\ln \left(1+\frac{1}{\omega}\right)$. We find that the CSST spectroscopic galaxy clustering survey can give $|\zeta|<10^{-2}$, or equivalently $|\omega|>\mathcal{O}(10^2)$ and  $|\dot{G}/G|<10^{-13}$, under the assumption $\zeta = 0$. These constraints are almost at the same order of magnitude compared to the joint constraints using the current cosmic microwave background (CMB), baryon acoustic oscillations (BAO), and Type Ia supernova (SN Ia) data, indicating that the CSST galaxy clustering spectroscopic survey would be powerful to constrain the BD theory and other modified gravity theories.
}

   \authorrunning{A. D. Chen et al.}            
   \titlerunning{Constraining BD theory with CSST spec-z survey}  

   \maketitle

%
%
\section{Introduction}           
\label{sect:intro}

The cosmic acceleration, which is discovered in 1998 (\citealt{Riess+etal+1998, Perlmutter+etal+1999}), is a great mystery of modern cosmology. To explain this phenomenon,  different kinds of dark energy (DE) model have been considered as it's origin, including the cosmological constant (CC) in the standard $\Lambda$CDM model. The mechanism of the cosmic acceleration  cannot be explained in the gravitation theory frame includeing only general relativity (GR) without DE, and that can also be seen as the incompleteness of GR. Thus, another alternative physical explanation is the modified gravity (MD), i.e., the modification of Einstein's general relativity theory.  Generally, the main difference of MD and DE is whether a theory violates the strong equivalence principle (SEP) or not. In other words, a DE model must obey the SEP, and a MD model does not comply with it (for more details, see, e.g. \citealt{Joyce+etal+2016}).  The Jordan-Fierz-Brans-Dicke theory (\citealt{Jordan+1949, Fierz+1956, Brans+Dicke+1961, Dicke+1962}, for a historical perspective, see \citealt{Brans+2014}, hereafter we call it Brans-Dicke (BD) theory for simplicity) is a typical example of modified scenario to GR. The BD theory is based on Mach's Principle, and it is the simplest case of the scalar-tensor theory of gravity and a natural alternative of GR. 

For decades, the astrophysical and astronomical observations give various constraints on the BD parameter $\omega$. According to  solar system data obtained from Cassini–Huygens mission, the estimation of $\omega$ is given as $\omega > 40000$ at $2\sigma$ ($\text{95.5\%}$) confident level (CL) (\citealt{Bertotti+etal+2003, Will+2006, Perivolaropoulos+2010}). For the modification of the evolution of our universe, some cosmological methods can also be used to test the BD theory.  There are several different ways of cosmological approach, such as cosmic microwave background (CMB), galaxy clustering, and Type Ia supernova (SN Ia). By using the combined CMB data and cosmic large-scale structure (LSS) measurements, including Wilkinson Microwave Anisotropy Probe (WMAP) five-year data, the Arcminute Cosmology Bolometer Array Receiver (ACBAR) 2007 data, the Cosmic Background Imager (CBI) polarization data, the Balloon Observations Of Millimetric Extragalactic Radiation and Geophysics (BOOMERanG) 2003 flight data, and the luminous red galaxy (LRG) survey of the Sloan Digital Sky Survey (SDSS) Data Release 4 (DR4), \citealt{Wu+etal+2010b} found $\omega > 97.8$ or $\omega < -120.0$ at $2\sigma$ CL. By using CMB data from WMAP, ACBAR, VSA, CBI, and galaxy power spectrum data from 2dF, \citet{Acquaviva+etal+2005} obtained $\omega > 120.0$ at $2\sigma$ CL. Based on the CMB temperature data from the {\it Planck} satellite and the nine-year polarization data from the WMAP, and  baryon acoustic oscillations (BAO) distance ratio data from the SDSS, \citet{Li+etal+2013} excluded the region of $\omega$ value in $-407.0<\omega<175.87$. Besides, using the combined data sets of CMB, BAO and SNIa, \citet{Li+etal+2015} excluded the region of $-9999.50<\omega<232.06$.

Since the LSS formation and evolution are tightly related to the properties of gravity, the Brans-Dicke theory and other modified gravity theories can be constrained by LSS related observations, such as BAO \citep[see e.g.][]{Eisenstein+2005}, weak gravitational lensing (WL; e.g., \citealt{Kaiser+1992, Kaiser+1998} and redshift-space distortion (RSD; \citealt{Jackson+1972, Kaiser+1987}). Some ongoing and planed telescopes are devoted to perform relevant measurements, e.g. SDSS \citep{Fukugita+etal+1996, York+etal+2000}, the Large Synoptic Survey Telescope (LSST, \citealt{Ivezic+etal+2008, Abell+etal+2009}) and \textit{Euclid} Space Telescope (\citealt{Laureijs+etal+2011}). The China Space Station Telescope (CSST) \citep{Zhan+2011, Zhan+2018, Cao+etal+2018, Gong+etal+2019} is also one of this kind of projects. The CSST, which is planed to launch around 2024, is a $2$m space telescope and in the same orbit with the China Manned Space Station. The telescope can carry out both photometric imaging and slitless-grating spectroscopic surveys simultaneously. The survey will cover $17,500$ $\text{deg}^{2}$ sky area with a field of view (FOV) of $1.1$ $\text{deg}^{2}$ in about 10 years. It has a wavelength coverage from near ultraviolet to near infrared with seven photometric and three spectroscopic bands. Compared to other surveys of next generation, the CSST has some advantages, e.g., large FOV, wide wavelength coverage, high image quality, and so on. It is expected to observe more than one hundred million galaxies from $z=0$ to 2 in its spectroscopic survey, that can precisely measure the evolution of the LSS and provide strong constraints on the modified gravity theories. In this study, we will generate mock data of the CSST spectroscopic galaxy survey, and explore its capability of constraining the BD theory.

\section{Basics of the Brans-Dicke Theory}

In the Brans-Dicke theory, the coupling of gravity and matter is still preserved, thus  the weak equivalence principle (WEP) is still unviolated, and all non-gravitational constants are unchanged. The action of BD theory in the usual frame is given by

\begin{equation}
\mathcal{S}=\frac{1}{16 \pi} \int d^{4} x \sqrt{-g}\left[-\phi R+\frac{\omega}{\phi} g^{\mu \nu} \nabla_{\mu} \phi \nabla_{\nu} \phi\right]+\mathcal{S}^{(m)},
\end{equation}
where the second term of the right hand side is the action of ordinary matter fields, which is given by $\mathcal{S}^{(m)}=\int d^{4} x \sqrt{-g} \mathcal{L}^{(m)}$,  the scalar field $\phi$ is Brans-Dicke field, and $\omega$ here is a dimensionless parameter.  The results of Cavendish type experiments require that  

\begin{equation}
\phi_{0}=\frac{2 \omega+4}{2 \omega+3} \frac{1}{G_{0}} .
\end{equation}
Here $\phi_{0}$ and $G_{0}$ are the present value of BD field and  the Newtonian gravitational constant. For convenience, we can then define a dimensionless field

\begin{equation}
\varphi=G \phi ,
\end{equation}
where $G$ is the Newtonian gravitational constant, which is variable in the BD theory, and $\varphi_{0}=G_{0} \phi_{0} $.  Thus, $G$ is related to a scalar field, and the value of the field is determined by all matter in the universe. In other words, the Mach principle is satisfied. 

In the limits of 

\begin{equation}
\omega \rightarrow \infty, \quad \varphi^{\prime} \rightarrow 0, \quad \varphi^{\prime \prime} \rightarrow 0,
\end{equation}
Brans-Dicke theory can be reduced to the General Relativity. To ensure the continuity of it's resulting range,  we introduce a new Brans-Dicke parameter \citep{Wu+etal+2010a}

\begin{equation}
\zeta=\ln \left(1+\frac{1}{\omega}\right).
\end{equation}
Then we have the limits $\zeta \rightarrow 0, \quad \varphi^{\prime} \rightarrow 0, \quad \varphi^{\prime \prime} \rightarrow 0$.
The BD theory modifies several aspects of $\Lambda$CDM model, especially the expansion history of the universe and the evolution of the LSS. In the BD theory, the modified Friedmann equation takes the form as \citep[see e.g.][]{Li+etal+2015}

\begin{equation}
H^{2}=\frac{\kappa}{3 \varphi} \rho+\frac{\omega}{6}\left(\frac{\dot{\varphi}}{\varphi}\right)^{2}-H \frac{\dot{\varphi}}{\varphi}.
\end{equation}
On the other hand, the calculation of the LSS evolution is much more complicated as shown in \citet{Wu+etal+2010a}, and we will discuss more details in the next section. In our Brans-Dicke cosmological model, for simplicity, we assume a flat universe with the cosmological constant as DE, which is supported by most of the current cosmological observations.

\section{Mock data and Model Constraints}

If the Brans-Dicke theory is considered as the alternative of GR,  the LSS evolution history and the interaction between galaxies will be different, and galaxy clustering is expected to present in different patterns. In this section, we will discuss the utility of galaxy clustering power spectrum measured by CSST spectroscopic survey as an approach to constrain the Brans-Dicke cosmological model.

\subsection{Mock Data of the CSST Galaxy Clustering Survey}

\begin{figure}
   \centering
   \includegraphics[width=0.8\textwidth, angle=0]{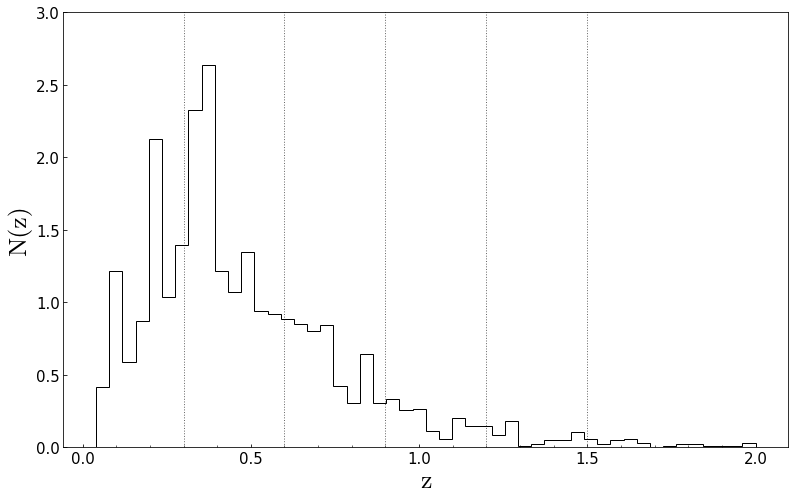}
   \caption{The mock galaxy redshift distribution of the CSST spectroscopic survey. The zCOSMOS catalog is adopted to derive the mock CSST spectroscopic galaxy catalog. The peak of the distribution is approximately at $z = 0.3 - 0.4$, and the distribution can extend to about $z=2$. The vertical lines denote the boundaries of redshift bins.}
   \label{fig1}
\end{figure}

The CSST will perform spectroscopic survey using slitless gratings with spectral resolution $R \gtrsim 200$, which contains three bands, i.e. GU, GV, and GI from 255 nm to 1000 nm.  For any point sources, their AB
magnitude $5\sigma$ limit is around 21 per resolution element or $\sim23$ for a band.
We adopt the mock catalog result of the CSST spectroscopic survey from \citealt{Gong+etal+2019}, which is derived from the zCOSMOS catalog (\citealt{Lilly+etal+2007,Lilly+etal+2009}). The zCOSMOS  is a redshift survey which performs the observations in the COSMOS field using the VIMOS spectrograph on the  Very Large Telescope (VLT). The zCOSMOS has a similar survey depth as the CSST spectroscopic survey, and its magnitude limit is $I_{\mathrm{AB}} \simeq 22.5$ and it covers the whole 1.7 $\text{deg}^{2}$ COSMOS field. Totally about 16,600 high-quality sources with reliable spectroscopic redshifts are selected, which corresponds to a galaxy surface density $\sim2.7$ arcmin$^{-2}$. The mock galaxy redshift distribution in the CSST spectroscopic survey is shown in Figure~\ref{fig1}. We can find that the distribution has a peak at $z=0.3-0.4$, and it can extend to $z\sim2$. As we discuss later, due to statistical requirement, we only use the sources at $z\le1.5$ when analyzing galaxy clustering.

\begin{figure}
   \centering
   \includegraphics[width=\textwidth, angle=0]{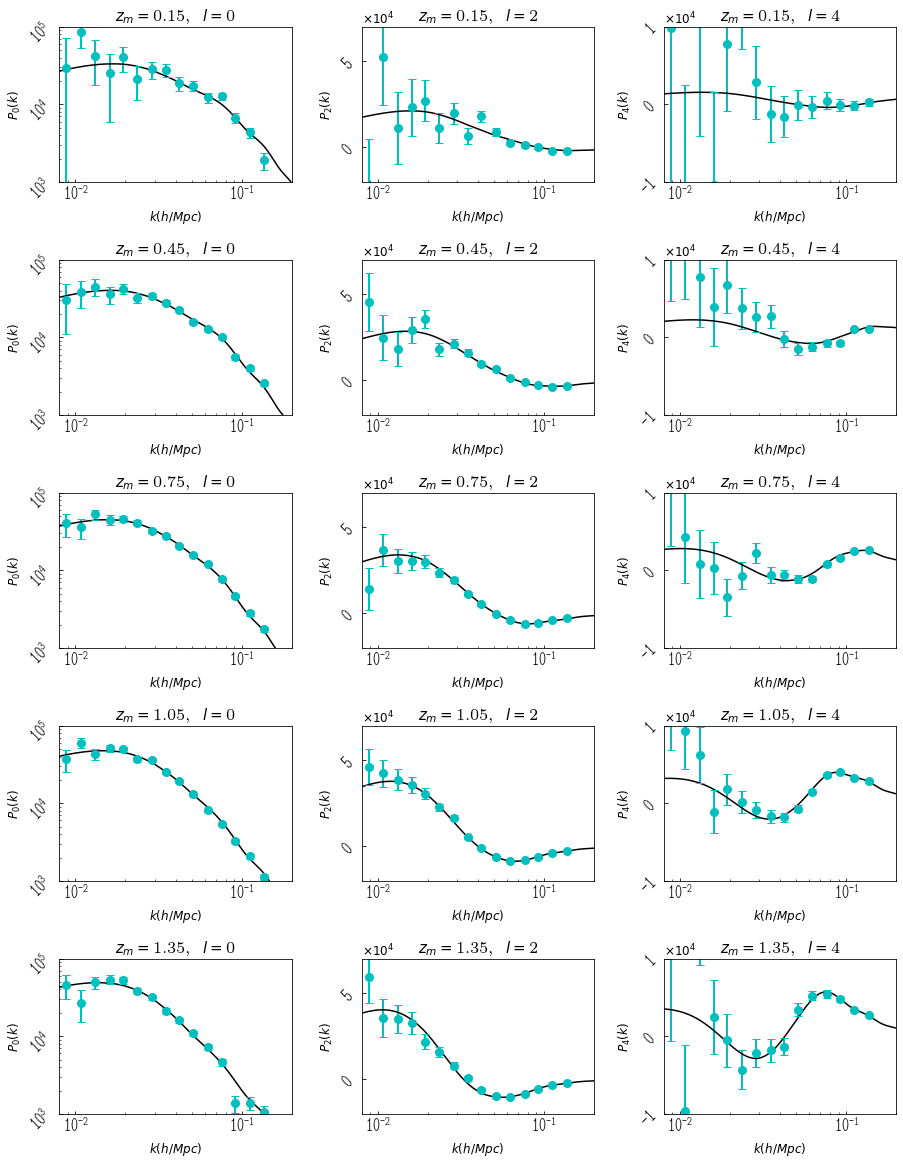}
   \caption{Galaxy mock multipole angular power spectra in the five spec-$z$ bins. The three columns from left to right are $P_{0}^g$, $P_{2}^g$ and $P_{4}^g$, respectively. The five rows show the five spec-$z$ bins (the $z_{m}$ denotes the central value of every single bin). To mimic statistical effect in the real measurements, the data points have been randomly shifted based on Gaussian distributions with the 1$\sigma$ values as the data errors.}
   \label{fig2}
\end{figure}

The CSST spectroscopic survey measures the galaxy clustering in redshift space, and we will discuss the measurements of galaxy correlation function or power spectrum considering the effects of redshift-space distortion. The redshift-space galaxy power spectrum can be expanded in Legendre polynomials (\citealt{Taylor+Hamilton+1996}) 

\begin{equation}
    P_{g}^{(\mathrm{s})}(k, \mu)=\sum_{\ell} P_{\ell}^{g}(k)     \mathcal{L}_{\ell}(\mu),
\end{equation}
where $P_{l}^{g}(k)$ is the multipole moments of the power spectrum, $\mu$ is the cosine of the angle between the line of sight and the direction of the wave number vector $\vec{k}$, $\mathcal{L}_{\ell}(\mu)$ is the Legendre polynomials which only include the first three non-zero orders $\ell=(0,2,4)$ in linear regime, and the superscript $s$ in the left side denotes the redshift space. 
The apparent redshift-space space galaxy power spectrum can be estimated as

\begin{equation}
    P_{g}^{(\mathrm{s})}\left(k^{\prime}, \mu^{\prime}\right)=P_{g}\left(k^{\prime}\right)\left(1+\beta \mu^{\prime 2}\right)^{2} \mathcal{D}\left(k^{\prime}, \mu^{\prime}\right),
\end{equation}
where $\mathcal{D}\left(k^{\prime}, \mu^{\prime}\right)$ is the damping term at small scales, which is given by

\begin{equation}
    \mathcal{D}\left(k^{\prime}, \mu^{\prime}\right)=\exp \left[-\left(k^{\prime} \mu^{\prime} \sigma_{\mathrm{D}}\right)^{2}\right],
\end{equation}
where $\sigma_{\mathrm{D}}^{2}=\sigma_{v}^{2}+\sigma_{R}^{2}$,  $\sigma_{v}$ is the velocity dispersion (\citealt{Scoccimarro+2004,Taruya+etal+2010}), and the relation between that and redshift is $\sigma_{v}=\sigma_{v_0} /(1+z)$, here we set $\sigma_{v_0}=7 \mathrm{Mpc} / h$ according to \citet{Blake+etal+2016} and \citet{Joudaki+etal+2018}. $\sigma_{R}=c \sigma_{z} / H(z)$ is so called smearing factor, which is the effect when the considered scale is less than the spectral resolution of spectroscopic surveys, and $\sigma_{z}=(1+z) \sigma_{z}^{0}$ (\citealt{Wang+etal+2009}). As a moderate consideration, we assume $\sigma_{z}^{0} = 0.002$. This value is based on the instrumental design of the CSST. Note that this damping term can only impact the power spectrum at small scales, and it would not affect the result significantly in the linear regime where we focus on in this work. $P_{g}\left(k^{\prime}\right)=b_{g}^{2} P_{\mathrm{m}}\left(k^{\prime}\right)$ is the apparent real-space galaxy power spectrum. $b_{g}$ is the galaxy bias, and $\beta = f/b_{g}$, where $f=d \ln D(a) / d \ln a$ is the growth rate, and we adopt the empirical fitting formula $f(a)\simeq\left[\Omega_{\mathrm{m}}(a)\right]^{0.55}$. The matter power spectrum $P_{m}$ can be calculated by using the $\texttt{CAMB}$ code (\citealt{Lewis+Bridle+2002}), and for the purpose of this paper, we adopt it's modified version (\citealt{Wu+etal+2010a}), which includes an implementation of cosmological model based on Brans-Dicke gravity theory. This code includes both modifications of background cosmology and full perturbation equations about structure growth in the BD theory.

In consideration of Alcock-Paczynski (AP) effect (\citealt{Alcock+Paczynski+1979}), one can write the galaxy multipole angular power spectra as

\begin{equation}
    P_{\ell}^{g}(k)=\frac{2 \ell+1}{2 \alpha_{\perp}^{2}     \alpha_{\|}} \int_{-1}^{1} \mathrm{~d} \mu     P_{g}^{(\mathrm{s})}\left(k^{\prime}, \mu^{\prime}\right)     \mathcal{L}_{\ell}(\mu).
\end{equation}
The scaling factors in the radial and transverse directions are given by

\begin{equation}
    \begin{array}{ll}
         & \alpha_{\|}=H^{\mathrm{fid}}(z) / H(z),\\
         & \alpha_{\perp}=D_{\mathrm{A}}(z) / D_{\mathrm{A}}^{\mathrm{fid}}(z).
    \end{array}
\end{equation}
The $H^{\mathrm{fid}}(z)$ and $D^{\mathrm{fid}}_{\mathrm{A}}(z)$ are Hubble parameter and angular diameter distance in fiducial cosmology, respectively. After we consider the effects of short noise and systematics, another two terms can be added to obtain the total multipole power spectra

\begin{equation}
    \widetilde{P}_{\ell}^{g, a}(k)=P_{\ell}^{g, a}(k)+\frac{1}{\bar{n}_{g}^{a}}+N_{\mathrm{sys}}^{g, a}.
\end{equation}
The superscript $a$ denotes different spectroscopic redshift bins. To analyze the redshift evolution effect and obtain more information, we divide the redshift range covered by the CSST spectroscopic survey into five bins from $z=0$ to 1.5 as shown in Figure~\ref{fig1}. $\bar{n}_{g}^{a}=f_{\mathrm{eff}}^{z_{s}, a} \bar{n}_{g, \mathrm{ori}}^{a}$ is the galaxy number density in a redshift bin. Considering that not all galaxies' redshifts in the CSST slitless spectroscopic survey can be well meadsured, an effective redshift-dependent fraction factor $f_{\text {eff }}^{z_{\mathrm{s}}}$ is included here. For simplicity, we assume

\begin{equation}
    f_{\text {eff }}^{z_{\mathrm{s}}} = \frac{f_{\text {eff }}^{z_{\mathrm{s}}, 0}}{1+z},
\end{equation}
we conservatively assume $f_{\text {eff }}^{z_{\mathrm{s}}, 0} = 0.5$, that only half of galaxies at $z=0$ can have well-measured redshift with spec-$z$ accuracy $\sim0.002$.
From the mock CSST spectroscopic catalog which is mentioned earlier, we obtain that $\bar{n}_{g, \text { ori }}^{a}=3.4 \times 10^{-2}, \quad 1.1 \times 10^{-2}, \quad 5.5 \times 10^{-3}, \quad 1.2 \times 10^{-3}, \quad 7.9 \times 10^{-5}$ $(\mathrm{Mpc} / h)^{-3}$ for the five spec-$z$ bins. The systematic term is included for the instrumental effects of the CSST slitless gratings, and we set it as a constant $N_{\mathrm{sys}}^{g, a} = \quad 5 \times 10^{4}$ $(\mathrm{Mpc} / h)^{-3}$, which can be seen as an average value for all of the spectroscopic redshift bins and scales. After that, we can estimate the error by

\begin{equation}
\sigma_{P_{\ell}^{g}}^{a}(k)=2 \pi \sqrt{\frac{1}{V_{\mathrm{S}}^{a} k^{2} \Delta k}} \widetilde{P}_{\ell}^{g, a}(k),
\end{equation}
where $V_{\mathrm{S}}^{a}$ is the survey volume in the $a$ th bin. The mock data of galaxy power spectra in the CSST spectroscopic survey are then obtained. 
The resulting mock galaxy power spectra in the five spec-$z$ bins are shown in Figure~\ref{fig2}. The three columns denote the first three non-vanished components of galaxy multipole angular power spectrum, i.e., $P_{0}^g$, $P_{2}^g$ and $P_{4}^g$. The five rows show the results in the five spec-$z$ bins, and the $z_{m}$ denotes the central value of every single bin. Here we consider the data points in the range of $k<0.2 \mathrm{~h} / \mathrm{Mpc}$, so that the nonlinear effect can be ignored.

\subsection{Fitting Method}

After obtaining the mock data of galaxy clustering measurements in the CSST spectroscopic survey, we will constrain the parameters in the model by using these data. 
We consider 11 free parameters in the model, including 1 Brans-Dicke parameter $\zeta$, 5 cosmological parameters, i.e. $\Omega_{c}h^{2}$, $\Omega_{b}h^{2}$, $H_{0}$, $A_s$ and $n_{s}$, and 5 galaxy bias parameters in the five spec-$z$ bins. $\Omega_{b}h^{2}$ and $\Omega_{c}h^{2}$ are the fraction of the total energy density of the universe contributed by baryonic matter and cold dark matter, respectively. $H_{0}$ is the Hubble constant, $A_s$ is the amplitude of primordial superhorizon power spectrum, and $n_{s}$ is the scalar spectral index.  After the fitting process, the derived parameters $\Omega_{\Lambda}$ and $\dot{G}/G$ can be obtained. Here $\Omega_{\Lambda}$ is the dark energy density, and $\dot{G}/G$ is the ratio of the time derivative of Newtonian gravitational constant and the constant itself. The velocity dispersion parameters are not considered as free parameters here, since the velocity dispersion mainly impacts the non-linear regime that cannot significantly affect our results. Note that the resulting mock data of galaxy mock power spectra in the CSST spectroscopic survey are derived under the assumption of $\zeta = 0$, namely the case of general relativity.

The $\chi^{2}$ statistic method is applied to fit our mock data, and the $\chi^{2}$ is defined as

\begin{equation}
\chi^{2}=\sum_{l,z_{m}} \left(\frac{P_{{l},{\mathrm{th}}}^{g}\left(z_{m}\right)-P_{{l},{\mathrm{mock}}}^{g}\left(z_{m}\right)}{\sigma_{\mathrm{mock}}}\right)^{2},
\end{equation}
where $P_{{l},{\mathrm{th}}}^{g}$ is the theoretical galaxy multipole angular power spectrum predicted from Brans-Dicke theory, $P_{{l},{\mathrm{mock}}}^{g}$ is also the galaxy clustering power spectrum but from the mock data, and $\sigma_{\mathrm{mock}}$ is the error. The summation is for all spec-$z$ bins and different $l$s. Then the likelihood function can be estimated by, 

\begin{equation}
\mathcal{L} \sim {\rm exp}(-\chi^{2} / 2).
\end{equation}
The prospective constraint on the Brans-Dicke model can be derived by performing the Markov Chain Monte Carlo (MCMC) method. In our analysis, the MCMC is implemented by the publicly code $\texttt{emcee}$ \citealt{Foreman-Mackey+etal+2013}, which is based on the  Goodman \& Weare’s affine-invariant ensemble sampler for MCMC. The fiducial value of input parameter and the range of flat prior are shown in Table~\ref{Tab:table1}.

\begin{table}[t]
\begin{center}
\caption[]{The fiducial values, flat priors, and fitting results of the free and derived parameters in our model.}\label{Tab:table1}
 \begin{tabular}{cccc}
  \hline\noalign{\smallskip}
Free parameter & Fiducial value &Flat prior& Fitting result (with 68\%  \& 95\% limits)\\
  \hline\noalign{\smallskip}
  $H_{0}$  &  67.5  & (50, 100)& $69.86_{-2.52-5.43}^{+2.86+7.55}$   \\
  $\Omega_{b}/h^2$  & 0.022  & (0.0, 0.5)& $0.0226_{-0.0033-0.0075}^{+0.0038+0.0097}$  \\
  $\Omega_{c}/h^2$  & 0.122  &  (0.0, 0.2)&$0.1248_{-0.0079-0.0157}^{+0.0094+0.0274}$  \\
  $A_{s}(\times 10^{-9})$& 2 & (0.1,3.9) & $2.06_{-0.71-1.64}^{+0.79+1.58}$ \\
  $n_{s}$&  0.965  & (0.9, 1)& $0.9648_{-0.0166-0.0427}^{+0.0161+0.0296}$  \\
  $\zeta$  &  0  & (-0.039, 0.039)& $-0.002_{-0.0036-0.0097}^{+0.0044+0.0081}$  \\
  \hline\noalign{\smallskip}
  Galaxy bias&1.15&(0,4)& $1.162_{-0.044-0.098}^{+0.043+0.095}$ \\
  & 1.45 &(0,4)& $1.453_{-0.031-0.081}^{+0.032+0.082}$ \\
  & 1.75 &(0,4)& $1.745_{-0.036-0.084}^{+0.033+0.080}$ \\
  & 2.05 &(0,4)& $2.043_{-0.036-0.080}^{+0.036+0.100}$ \\
  & 2.35 &(0,4)& $2.339_{-0.044-0.096}^{+0.040+0.093}$ \\
  
  \hline\noalign{\smallskip}
  Derived Parameter &&& Fitting result (with 68\%  \& 95\% limits)\\
  \hline\noalign{\smallskip}
  $\Omega_{\Lambda}$ & &&$0.6794_{-0.0080-0.00355}^{+0.0078+0.0292}$ \\
  $\dot{G} / G(\times 10^{-13})$ & & &$0.2238_{-0.3985-0.8369}^{+0.4975+1.1464}$ \\
  \hline\noalign{\smallskip}
\end{tabular}
\end{center}
\end{table}

\section{Results}

\begin{figure}[t]
 \centering
 \includegraphics[width=\textwidth, angle=0]{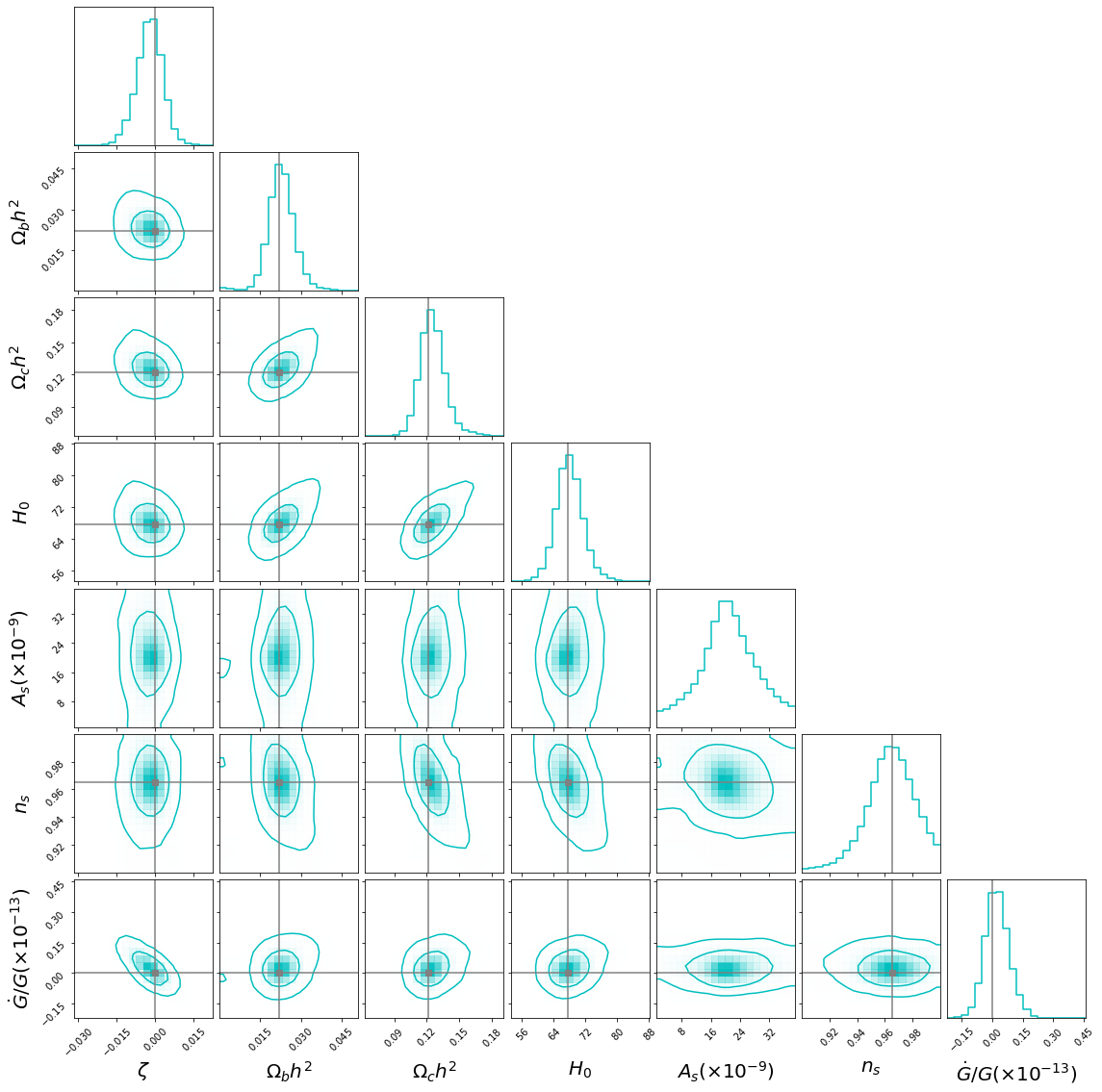}
 \caption{The 2-D contour maps (68\% and 95\% C.L.) and 1-D histograms of the posterior probability distributions of the free and derived cosmological parameters in our BD model. The fiducial values of the parameters are also shown in gray lines.}
 \label{fig3}
 \end{figure}

\begin{figure}[t]
 \centering
 \includegraphics[width=\textwidth, angle=0]{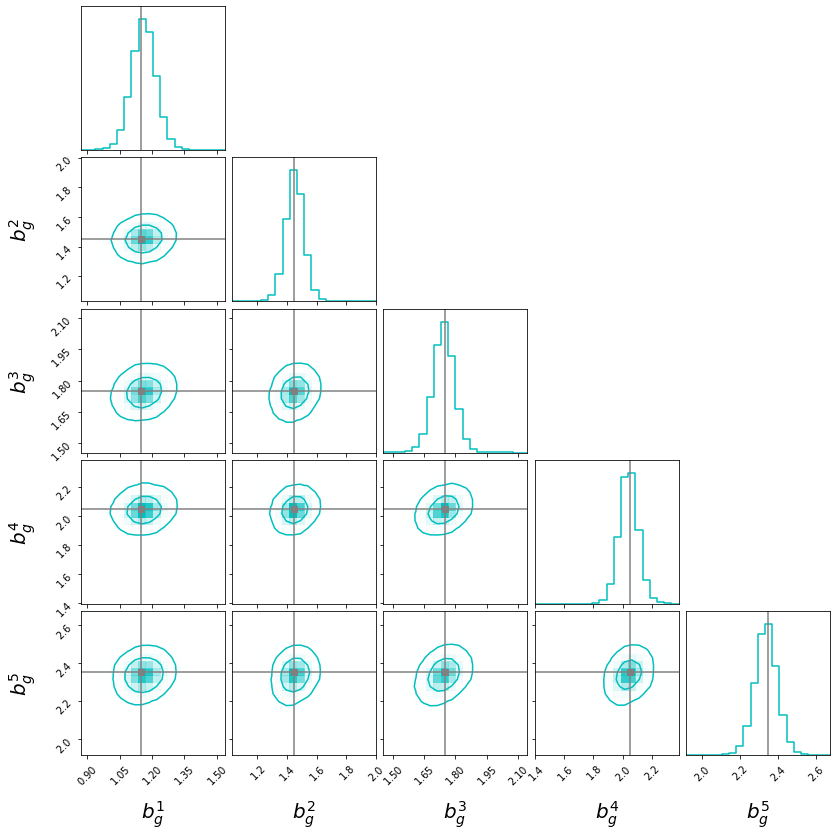}
 \caption{The 2-D contour maps (68\% and 95\% C.L.) and 1-D histograms of the posterior probability distributions of the galaxy bias in the 5 spec-$z$ bins. The fiducial values of the parameters are also shown in gray lines.}
 \label{fig4}
 \end{figure}

After the MCMC fitting process, we have the constraint results of the free parameters in our Brans-Dicke model.
The best-fits and 1$\sigma$ and 2$\sigma$ errors of the free and derived parameters are shown in Table~\ref{Tab:table1}. The 2-D projected contour maps and 1-D histograms of the posterior probability distributions of our BD cosmological parameters and galaxy bias are shown in Figure~\ref{fig3} and Figure~\ref{fig4}, respectively.
 
We can find that the Brans-Dicke parameters $\zeta$ has been constrained as

\begin{equation}
-0.56 \times 10^{-2}<\zeta<0.24 \times 10^{-2}(68 \%\ {\rm C.L.}),
\end{equation}
\begin{equation}
-1.17 \times 10^{-2}<\zeta<0.61 \times 10^{-2}(95 \%\ {\rm C.L.}).
\end{equation}
or the results can be converted to $\omega$ as

\begin{equation}
(\omega >416.17 ) \cup (\omega<-179.07)(68\%\ {\rm C.L.}),
\end{equation}
\begin{equation}
(\omega > 163.43) \cup (\omega<-85.97)(95\%\ {\rm C.L.}).
\end{equation}
Besides, using the relation $\dot{G} / G \equiv-\dot{\varphi} / \varphi$, the rate of change of the Newtonian gravitational constant can also be found as

\begin{equation}
-0.1747 \times 10^{-13}<\dot{G} / G<0.7213 \times 10^{-13}(68 \%\ {\rm C.L.}),
\end{equation}
\begin{equation}
-0.8369 \times 10^{-13}<\dot{G} / G<1.3702 \times 10^{-13}(95 \%\ {\rm C.L.}).
\end{equation}

Compared to previous results, e.g. the joint constraints by using CMB+BAO+SN Ia data \citep{Li+etal+2015}, 
our constraints on $\zeta$, $\omega$ or $\dot{G}/G$ are at the same order of magnitude and comparable to the previous ones. Note that here we only consider the galaxy clustering measurement in the CSST spectroscopic survey, the results can be further improved by including other CSST observations, such as weak and strong gravitational lensing, 2-D angular galaxy clustering, galaxy clusters, and so on.

Besides, the constraints on the other cosmological parameters are also strong, especially for $\Omega_{c}h^{2}$, $\Omega_{b}h^{2}$ and $H_{0}$, which are basically consistent with previous results in a $\Lambda$CDM universe with GR assumed \citep{Gong+etal+2019}. In addition to the cosmological parameters, we also simultaneously constrain the galaxy bias parameters in different spec-$z$ bins as shown in Figure~\ref{fig4}. We can find that the CSST spectroscopic galaxy clustering measurement can provide effective constraints on galaxy bias with an uncertainty around $\pm0.1$ (68\% C.L.), which is also in a good agreement with the results given in \citet{Gong+etal+2019}. The best-fits of galaxy biases also can be correctly derived.

\section{Summary and Conclusion}

In this paper, we test the ability of the CSST galaxy clustering spectroscopic survey in constraining the cosmology in the Brans-Dicke framework. We generate mock catalog based on the zCOSMOS survey considering CSST instrumental and observational effects. The galaxy clustering multipole power spectra in different spec-$z$ bins are calculated based on a modified $\tt CAMB$ code considering BD theory and redshift-space distortion effect. The parameter $\zeta$ or $\omega$ of BD theory and other cosmological parameters are considered in the model. The galaxy bias parameters in different spec-$z$ bins are also included and simultaneously constrained in the fitting process. The MCMC have been performed for fitting all of eleven free parameters, and the contour maps and 1-D probability distribution functions of the parameters are obtained. Finally, we get the constraint intervals of $\zeta$, $\omega$ and $\dot G/G$. We find that the CSST spectroscopic galaxy survey can put strong constraints on the BD theory with $|\zeta|<10^{-2}$, $|\omega|>\mathcal{O}(10^2)$ and  $|\dot{G}/G|<10^{-13}$, that is at the same order of magnitude compared to the constraints using current joint datasets of cosmological observations. The results can be further improved by including other CSST measurements, which indicates that the CSST can provide powerful surveys to constrain the BD theory and other modified gravity theories.

\begin{acknowledgements}
A.D.C. and Y.G. acknowledge the support of MOST-2018YFE0120800, 2020SKA0110402, NSFC-11822305, NSFC-11773031, NSFC-11633004, and CAS Interdisciplinary Innovation Team. F.Q.W. acknowledges the Chinese Academy of Sciences (CAS) instrument grant ZDKYYQ20200008, the CAS Strategic Priority Research Program XDA15020200. Y.G.W. acknowledges National Science Foundation of China (Grant No. 11773034, and 11633004), the Chinese Academy of Sciences (CAS) Strategic Priority Research Program XDA15020200 and and the CAS Interdisciplinary Innovation Team (JCTD- 2019-05). X.L.C. acknowledges the support of the National Natural Science Foundation of China through grant No. 11473044, 11973047, and the Chinese Academy of Science grants QYZDJ-SSW-SLH017, XDB 23040100.
\end{acknowledgements}

\label{lastpage}

\end{document}